\documentclass[pra,twocolumn,superscriptaddress,amssymb]{revtex4}
\usepackage[dvipdfmx]{graphicx}
\usepackage{amsmath}
\usepackage{pdfpages, epsfig}
\DeclareMathOperator{\sech}{sech}

\begin{document}
\title{Core filling and snaking instability of dark solitons in spin-imbalanced superfluid Fermi gases}

\author{Matthew D. \surname{Reichl}}
	\affiliation{Laboratory of Atomic and Solid State Physics, Cornell University, Ithaca, New York 14853, USA}
	
\author{Erich J. \surname{Mueller}}
	\affiliation{Laboratory of Atomic and Solid State Physics, Cornell University, Ithaca, New York 14853, USA}

\date{\today}

\begin{abstract}        
 We use the time-dependent Bogoliubov de Gennes equations to study dark solitons in three-dimensional spin-imbalanced superfluid Fermi gases. We explore how the shape and dynamics of dark solitons are altered by the presence of excess unpaired spins which fill their low-density core. The unpaired particles broaden the solitons and suppress the transverse snake instability. We discuss ways of observing these phenomena in cold atom experiments.
\end{abstract}
\maketitle

\section{Introduction}
Ultracold atoms have become the best platform for studying collective nonlinear phenomena such as dark solitons. Dark solitons are persistent nonlinear collective excitations in which the density is reduced in a plane. They have been studied in a number of physical settings including atomic Bose-Einstein Condensates (BECs) \cite{frantzeskakis2010} and superfluid Fermi gases of spin-1/2 atoms \cite{antezza2007, liao2011, scott2011, yefsah2013}. They are ubiquitous in quenches \cite{lamporesi2013, donadello2014} and can be engineered through phase imprinting protocols \cite{burger1999, denschlag2000, anderson2001, becker2008, yefsah2013, ku2014, ku2016}.  Previous experimental \cite{tikhonenko1996, anderson2001, dutton2001, donadello2014, ku2014, ku2016} and theoretical work \cite{zakharov1974, jones1986, kuznetsov1988, mamaev1996, muryshev1999, feder2000, musslimani2001, brand2002, kamchatnov2008, brazhnyi2011, cetoli2013, mateo2014, bulgac2014, reichl2013, lombardi2016} has discovered that these dark solitons are dynamically unstable to a ``snaking" instability transverse to the plane of the soliton in both BECs and Fermi gases. In this paper, we theoretically study the dynamics of dark solitons in superfluid Fermi gases in which there is an imbalance between the number of up and down spins in the system. We find that the snaking instability is suppressed by the presence of excess spins which reside within the density depleted plane--or core--of the soliton. 

Previous work \cite{cetoli2013} has studied the snaking instability in spin-balanced Fermi gases using similar approaches as our paper. More recently, the authors of Refs.~ \cite{lombardi2016b, lombardi2016} applied an effective field theoretic approach \cite{klimin2015} to studying core filling and snaking instabilities of dark solitons in imbalanced Fermi gases. In this paper, we take a more microscopic approach and model the Fermi gas using the Bogoliubov de Gennes (BdG) equations. Our work also extends recent simulations of the stability of one-dimensional soliton trains \cite{dutta2016} which suggest that excess spin can stabilize dynamical instabilities of dark solitons.

The BdG theory captures the phenomenology of the BEC-BCS crossover \cite{zwerger2011}: at strong attractive interactions the fermions form tightly bound bosonic pairs which condense into a BEC, while at weak interactions the fermions form Cooper pairs which form a neutral analog to a BCS superconductor. Here we study the unitary gas which lies between these two limits. We caution that the BdG theory is a mean-field theory which only approximately models strong correlation physics in the unitary gas. However, the BdG equations has been successfully utilized in previous studies of dark soliton profiles and dynamics in the unitary gas \cite{antezza2007, liao2011, scott2011, spuntarelli2011, scott2012} and appears to be semi-quantitative.

In Sec.~\ref{secstat} we discuss the BdG model and find stationary dark solitons in the presence of imbalance. In Sec.~\ref{secdyn} we apply time-dependent BdG equations to simulate the snaking instability. In Sec.~\ref{secdis} we discuss how our results might be observed in cold atom experiments.

\section{Stationary Dark Solitons} \label{secstat}

\subsection{Model} \label{secmodel}

We consider the following Hamiltonian which describes spin-imbalanced spin-1/2 fermions with short-range attractive interactions
\begin{equation}
\begin{split}
\hat{H}= & \int{d^3\vec{x} \Big[ \sum_{\sigma = \uparrow, \downarrow} \Psi^\dagger_\sigma(\vec{x}) (-\frac{ \hbar^2 \nabla^2}{2m}} - \mu_\sigma) \Psi_\sigma(\vec{x}) \\
 & -g \Psi^\dagger_\uparrow(\vec{x})\Psi^\dagger_\downarrow(\vec{x}) \Psi_\downarrow(\vec{x})  \Psi_\uparrow(\vec{x}) \Big]
\end{split}
\end{equation}
Here $\mu_\sigma$ is the chemical potential for spin component $\sigma$ and $g$ is the bare interaction strength, which is related to the s-wave scattering length $a_s$ by 
\begin{equation} \label{renormeq}
1/g= -m/(4\pi \hbar^2 a_s) + \frac{1}{V}  \sum_k 1/(2 \epsilon_k)
\end{equation}
where $V$ is the volume of the system and $\epsilon_k=\frac{\hbar^2 k^2}{2m}$. The sum $  \sum_k 1/(2 \epsilon_k)$ is taken over the discrete momenta determined by the grid spacing of our numerics and comes with a momentum cutoff of $k_{c_\nu} = N_{\rm{grid}}  \pi/ L_\nu $ (with high energy cutoff $E_c = \hbar^2 k_c^2 /2m$) where $L_\nu$ is the length of the system along the direction $\nu$ and $N_{\rm{grid}}$ is the number of grid points along that direction.  In this paper we focus our attention on the unitary limit $a_s \rightarrow \infty$. 

At zero temperature, up and down spin atoms combine into Cooper pairs and condense to form a superfluid. We rewrite $\hat{H}$ in terms of the Cooper pair field $\Delta(\vec{x}) = g \langle \Psi_\uparrow(\vec{x})  \Psi_\downarrow(\vec{x}) \rangle $ and neglect quadratic fluctuations. This gives the following mean-field BdG Hamiltonian \cite{de1966}
\begin{equation} 
\begin{split}
& \hat{H}_{\rm{BdG}} = \\ 
 & \int{ \! \!d^3\vec{x}} \left( \begin{array}{ccc}
\Psi_\uparrow(\vec{x})  \\
\Psi^\dagger_\downarrow(\vec{x}) \end{array} \right)^\dagger  \! \! \left( \begin{array}{ccc}
-\frac{\hbar^2 \nabla^2}{2m} - \mu_\uparrow & \Delta(\vec{x})  \\
 \Delta^*(\vec{x}) & \frac{\hbar^2 \nabla^2}{2m} + \mu_\downarrow \end{array} \right) \! \! \left( \begin{array}{ccc}
\Psi_\uparrow(\vec{x})  \\
\Psi^\dagger_\downarrow(\vec{x}) \end{array} \right) 
\end{split}
\end{equation}

$H_{\rm{BdG}}$ is diagonalized through a Bogoliubov transformation
\begin{equation}
 \left( \begin{array}{ccc}
\Psi_\uparrow(\vec{x})  \\
\Psi^\dagger_\downarrow(\vec{x}) \end{array} \right) = \sum_n  \left( \begin{array}{ccc}
u_n(\vec{x}) & -v^*_n(\vec{x})  \\
v_n(\vec{x}) & u^*_n(\vec{x}) \end{array} \right)  \left( \begin{array}{ccc}
\gamma_{\uparrow, n}  \\
\gamma^\dagger_{\downarrow, n} \end{array} \right)
\end{equation}
where $\gamma^\dagger_{\sigma, n}$ is the creation operator for a Bogoliobov excitation of energy $E_{\sigma, n}= E_n \pm h$ where  $E_n$ are the positive eigenvalues of the equation
\begin{equation} \label{bdgeq}
 \left( \begin{array}{ccc}
-\frac{\hbar^2 \nabla^2}{2m} - \mu & \Delta(\vec{x})  \\
 \Delta^*(\vec{x}) & \frac{\hbar^2 \nabla^2}{2m} + \mu \end{array} \right)  \left( \begin{array}{ccc}
u_n(\vec{x})  \\
v_n(\vec{x}) \end{array} \right) = E_n \left( \begin{array}{ccc}
u_n(\vec{x})  \\
v_n(\vec{x}) \end{array} \right)
\end{equation}
and where $h$ and $\mu$ are given by $h = \frac{1}{2}(\mu_{\uparrow} - \mu_{\downarrow})$ and $\mu=  \frac{1}{2}(\mu_{\uparrow} + \mu_{\downarrow})$. At zero temperature, $\Delta(\vec{x})$ is expressed in terms of $u$'s and $v$'s as 
\begin{equation} \label{deleq}
\Delta(\vec{x}) = g \sum_{E_n>0} u_n(\vec{x}) v^*_n(\vec{x}) (1 - \Theta(-E_{\uparrow,n}) - \Theta(-E_{\downarrow, n}))
\end{equation}
where $\Theta(x)$ is the unit step function. The density $n_\sigma(\vec{x}) $ of fermions with spin $\sigma$  is given by 
\begin{equation} \label{denseq}
\begin{split}
& n_\sigma  =   \langle \Psi^\dagger_\sigma(\vec{x}) \Psi_\sigma(\vec{x}) \rangle \\
& = \sum_{E_n>0} \big[ |u_n(\vec{x})|^2 \Theta(-E_{\sigma, n}) +  |v_n(\vec{x})|^2 (1-  \Theta(-E_{-\sigma, n})) \big]
\end{split}
\end{equation}

\subsection{Numerical results} \label{secresults}

We numerically solve the coupled equations (\ref{bdgeq}) and (\ref{deleq}) using an iterative procedure. We first choose an ansatz pair field $\Delta(\vec{x})= \Delta \tanh(x/\xi)$ corresponding to a planar dark soliton fixed at $x=0$. $\xi$ parametrizes the width of the soliton core and is generally chosen in the ansatz to be $\xi \approx k_F$, where $k_F \equiv (3 \pi ^2 n_o)^{1/3}$  is the Fermi wavevector and $n_o$ is the density far from the core of the soliton. We then solve Eq.~(\ref{bdgeq}) and calculate a new $\Delta$ from Eq.(\ref{deleq}). This process is repeated until $\Delta$ converges to a stationary solution. In all the calculations presented in this paper we check that $\Delta$ converges to same stationary solution after small changes to the initial ansatz.

For simplicity we consider a system in a rectangular box geometry with dimensions $L_x \times L_\perp \times L_\perp$. We impose periodic boundary conditions in the $y$ and $z$ (perpendicular) directions, and in the $x$-direction we impose the conditions: $u_n(x+L_x)= u_n(x)$ and $v_n(x+L_x)= - v_n(x)$. This boundary condition in the $x$-direction ensures that $\Delta(x+L_x) = -\Delta(x)$, which is consistent with the profile of a single dark soliton in a finite size box. Because of the homogeneity of the stationary soliton in the perpendicular directions, the solution to  Eq.~(\ref{bdgeq}) can be expressed in the form $u_n(\vec{x})= u_{m, k_y, k_z}(x) \exp(i k_y y+ i k_z z)$, $v_n(\vec{x})= v_{m, k_y, k_z}(x) \exp(i k_y y+ i k_z z)$. This effectively reduces the three dimensional problem to a series of one-dimensional problems for each $k_y$ and $k_z$, and substantially speeds up the calculation. 

\begin{figure*}
\hbox{\hspace{-0.0em}
\includegraphics[width=1.0\textwidth]{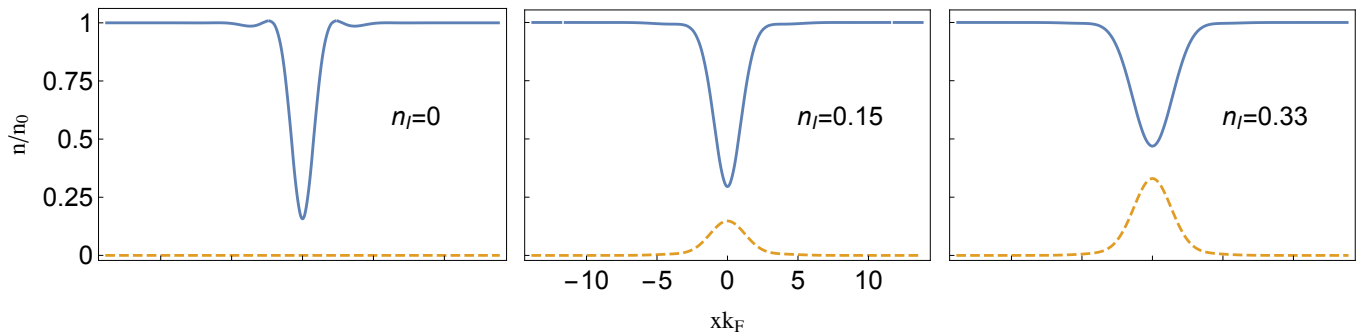}}
\caption{(Color online) Density profiles of a dark soliton at different relative spin imbalances $n_I=0, 0.15, 0.33$  (Eq.~\ref{nieq}) . The solid blue curves show the total density $n_\uparrow+ n_\downarrow$ and the dashed orange curves show the density difference $n_\uparrow - n_\downarrow$. The densities are plotted as a function of $x$ after integrating over the $y$ and $z$ directions and normalizing by the asymptotic density $n_o$} 
\label{denspic}
\end{figure*}

Fig.~\ref{denspic} shows the total density $n_\uparrow+n_\downarrow$  (solid blue curve) and density difference $n_\uparrow-n_\downarrow$ (dashed orange curve) for stationary dark soliton solutions in the presence of spin imbalance. The densities are plotted as functions of $x$ after integrating over the $y$ and $z$ directions. In these calculations the dimensions of the box are set to $L_x k_f \approx 28$ and $L_\perp k_f \approx 23$, and we use 60 grid points along the $x$-direction and 50 $k$-space points in both perpendicular directions.  These numerical parameters correspond to an energy cutoff   $E_c \approx 100 E_F$ in the sum in Eq.~(\ref{renormeq}). Our results are unchanged by using more grid points.  

 We characterize the spin imbalance using the relative spin imbalance $n_I$:
\begin{equation} \label{nieq}
n_I = \frac{n_{\uparrow}(x=0)- n_{\downarrow}(x=0)}{n_o}
\end{equation}
where as before $n_o$ is the total density far from the core. Fig.~\ref{denspic} shows a range of imbalances from $n_I=0$ to $n_I=0.33$. As the imbalance is increased, the soliton core (visually represented in the figure as the dip in the total density at $x=0$) fills with excess up spins and widens. This is consistent with previous calculations using different methods \cite{lombardi2016b, lombardi2016} and is expected given simple energetic considerations: the most energetically favorable place to store excess unpaired spins is at the core of the soliton where $\Delta =0$ and hence no Cooper pairs need to be broken. On a more microscopic level, the soliton supports a band of midgap Andreev states which are bound to the core of the soliton \cite{antezza2007} which are filled by excess spins after tuning the chemical potential bias $h$ away from 0.

\section{Snaking Instability}\label{secdyn}

\begin{figure}
\hbox{\hspace{-1.0em}
\includegraphics[width=0.5\textwidth]{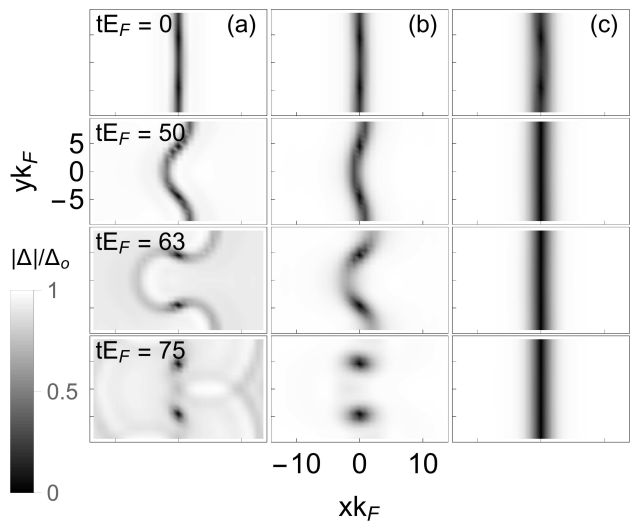}}
\caption{Dynamics of the snaking instability of dark solitons in the presence of excess spins. Dark colors are regions of small $|\Delta(x,y, z=0)|/\Delta_o$, where $\Delta_o$ is the value of $\Delta$ far from the soliton core at time $t=0$. The relative spin imbalances $n_I$ at the core of the soliton (Eq.~\ref{nieq})  are $n_I=0, 0.18, 0.40$ for columns (a), (b), (c), respectively. The transverse length is $L_\perp k_F \approx 18 $. } 
\label{timepic}
\end{figure}

In this section we discuss time dependent simulations of the snaking instability of dark solitons in the presence of spin imbalance. We find that the instability proceeds slower or, for sufficiently high imbalance, is completely suppressed by the presence of excess spins in the core of the soliton.

We numerically solve the following time-dependent BdG equations:
\begin{equation} \label{tbdgeq}
 \left( \begin{array}{ccc}
-\frac{\hbar^2 \nabla^2}{2m} - \mu & \Delta(\vec{x}, t)  \\
 \Delta^*(\vec{x},t) & \frac{\hbar^2 \nabla^2}{2m} + \mu \end{array} \right)  \left( \begin{array}{ccc}
u_n(\vec{x},t)  \\
v_n(\vec{x},t) \end{array} \right) = i \hbar \frac{\partial}{\partial t} \left( \begin{array}{ccc}
u_n(\vec{x},t)  \\
v_n(\vec{x},t) \end{array} \right)
\end{equation}
where 

\begin{equation} \label{tddel}
\Delta(\vec{x}, t) = g \sum_{E_n>0} u_n(\vec{x}, t) v^*_n(\vec{x}, t) (1 - \Theta(-E_{\uparrow,n}) - \Theta(-E_{\downarrow, n}))
\end{equation}

The initial set of $u_n(\vec{x}, 0)$ and $v_n(\vec{x}, 0)$ and the $E_n$'s in Eq.~(\ref{tddel}) are stationary solutions of Eqs. (\ref{bdgeq}) and (\ref{deleq}). For simplicity we again consider a system at unitarity ($a_s = \infty$) in a rectangular box geometry of dimensions $L_x \times L_\perp \times L_\perp$, and we use the same boundary conditions as in Sec.~\ref{secresults}. We assume homogeneity along the $z$-direction and express the $u_n$'s and $v_n$'s in the form $u_n(\vec{x},t)= u_{m, k_z} (x,y,t) \exp(i k_z z)$ and $v_n(\vec{x},t)= v_{m, k_z} (x,y,t) \exp(i k_z z)$. We use approximately 1000 grid points in the $x-y$ plane and 25 $k_z$ points which corresponds to an energy cutoff $E_c \approx 40 E_F$.

In all the simulations described here, we first perturb the stationary $\Delta(x)$ by adding a small term which seeds a snaking instability along the y-direction:
\begin{equation}
\Delta(\vec{x}, 0) = \Delta(\vec{x}) + \epsilon \sech(x)\sin \frac{2 \pi}{L_y} (y-L_y/4) 
\end{equation}
where $\epsilon \approx 0.1 \Delta_o$ and $\Delta_o$ is the value of $\Delta$ far from the soliton core.
 We then discretize time and evolve the set of $u_n$'s and $v_n$'s forward by one time step using a split step method with $\Delta$ calculated from the current time-step. After finding the new $u_n$'s and $v_n$'s, we calculate $\Delta(t)$ at the next time-step using Eq.~(\ref{tddel}). Details of the split step method are discussed in the appendix.
 
 Figure~\ref{timepic} shows the dynamics of a dark soliton for three different relative imbalances $n_I= 0, 0.18, 0.4$ (columns (a), (b), (c), respectively). The figures show graphs of $|\Delta(x,y, z=0)|/\Delta_o$ at different times. In these graphs we have $L_x k_F \approx 28$ and $L_\perp  k_F \approx 18$.
At zero imbalance there is clearly a snaking instability whose rate is consistent with similar calculations in other work \cite{cetoli2013}. The plane of the soliton buckles and eventually breaks leaving behind two vortex cores. However at $n_I=0.18$, the instability occurs at a slower rate and finally at $n_I=0.4$ the instability is completely suppressed. We have run these simulations up to times of $t= 150/E_F$, finding no sign of a snaking instability for large imbalances.

\begin{figure} 
\hbox{\hspace{1.0em}
\includegraphics[width=0.45\textwidth]{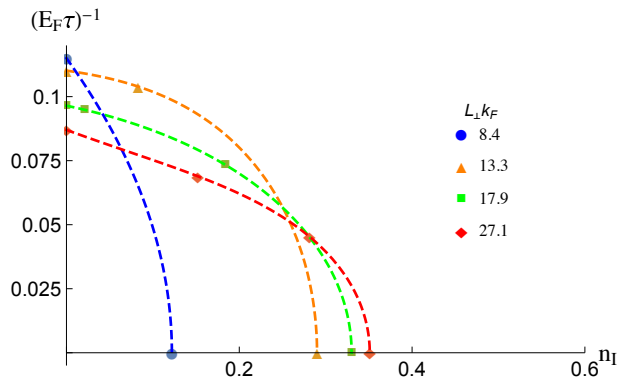}}
\caption{(Color online) Time scale $\tau$ for the decay of a dark soliton via a transverse snaking instability as a function of the relative spin imbalance $n_I$ (Eq.~(\ref{nieq})) at the core of the soliton . The different colors and shapes represent different transverse lengths $L_\perp$. The dashed curves show guides for the eye.} 
\label{timescales}
\end{figure}

Figure~\ref{timescales} shows time scales for the snaking instability at different imbalances and different transverse dimensions $L_\perp$. Each time scale $\tau$ was calculated by first extracting the position of the core $x_{\textrm{core}}(t)$ along the $y=0$ line, and fitting it to a exponential function $x_{\textrm{core}}(t) \sim \exp(t/\tau)$. At zero imbalance the time scale gets larger as $L_\perp$ increases. This trend is somewhat counter-intuitive, but can be understood by noting that the unstable mode's wavelength grows as $L_\perp$ increases. The instability connects with the Goldstone mode, and hence its frequency $\omega= i \hbar/ \tau$ vanishes as $L_\perp \to \infty$. Similar results were seen in Ref.~\cite{cetoli2013}. For larger $L_\perp$ (beyond those shown in this figure) additional decay modes appear. For sufficiently small $L_\perp$ ($\frac{\hbar^2}{m L_\perp^2} \sim \mu$) the rate will again decrease as the system becomes quasi-one dimensional. For small enough $L_\perp$ the soliton is stable, even without imbalance.

Once the soliton core is filled with excess spins the rate of the snaking instability becomes slower and is eventually suppressed altogether (up to times of at least $t E_F = 150$). At smaller $L_\perp$ the snaking instability is suppressed for smaller values of spin imbalance.

\section{Discussion} \label{secdis}

In this paper we have studied the dynamics of a dark soliton in an imbalanced Fermionic superfluid. We have found that the snaking instability of the soliton is suppressed by the presence of excess spins which reside at the low density core of the soliton. Although we have focused on the unitary limit, we expect these phenomena to persist in the BEC regime ($a_s \gg 0$) and BCS regime ($a_s \ll 0$). 
The behavior in the BEC regime should be similar to that of a two-component BEC: the paired superfluid maps onto one component, while the unpaired fermions in the core can crudely be modeled by a second component. Indeed, calculations \cite{musslimani2001, brazhnyi2011} and experiments \cite{anderson2001} of solitons in two-component BECs observe a suppression of the snaking instability.


We feel that experimentally observing the physics presented in this paper is feasible given existing tools. In a trapped imbalanced Fermi gas at equilibrium, excess spins reside along the edge of the trap \cite{olsen2015}. One naive idea is therefore to phase imprint a soliton onto the system and allow for the excess spins to diffuse from the edge of the trap into the core of the soliton. Unfortunately, the time scales for this process are prohibitively long. Instead we suggest first using a laser to create a potential barrier across the center of the trap and separating the imbalanced superfluid into two disjoint regions.  This geometry was produced in Ref.~\cite{valtolina2015}. Excess spins will then reside at the center of the trap between the two superfluid halves. Phase imprinting should then result in a soliton whose core is at the location of the excess spins. Varying the shape and dynamics of the applied potential barrier should allow experimentalists to control the relative imbalance $n_I$. One can image the soliton in time of flight after a ramp to the BEC limit as done in Refs.~\cite{yefsah2013, ku2014, ku2016}.

\section{Acknowledgements}
This material was based upon work supported by the National Science Foundation Grant No. PHY-1508300 and the ARO-MURI Non-Equilibrium Many-Body Dynamics Grant W9111NF-14-1-0003.
\bibliography{imbsolpaper}

\section{Appendix}
In this appendix we discuss the details of the split step method used to perform the time-dependent BdG simulations discussed in Sec.~\ref{secdyn}. Assuming homogeneity in the z-direction, the time-dependent BdG equations are given by 

\begin{equation} 
\widetilde{H} \left( \begin{array}{ccc}
u_{n, k_z}(x, y, t)  \\
v_{n, k_z}(x,y, t) \end{array} \right) = i \hbar \frac{\partial}{\partial t} \left( \begin{array}{ccc}
u_{n, k_z}(x,y, t)  \\
v_{n, k_z}(x,y, t) \end{array} \right)
\end{equation}

 where 
 \begin{widetext}
 \begin{equation} \label{hamileq}
  \widetilde{H}= \left( \begin{array}{ccc}
-\frac{\hbar^2}{2m}(  \frac{\partial^2}{\partial x^2}+ \frac{\partial^2}{\partial y^2} - k_z^2)- \mu & \Delta(x,y, t)  \\
 \Delta^*(x,y,t) & \frac{\hbar^2}{2m}(\frac{\partial^2}{\partial x^2}+ \frac{\partial^2}{\partial y^2} - k_z^2)  + \mu \end{array} \right)
 \end{equation}
 \end{widetext}
 and $ \Delta(x,y, t)$ is given by Eq.~(\ref{tddel}) in the main text. We approximate the operator $\exp [- i  \widetilde{H} \delta t ]$ which evolves the $u$'s and $v$'s forward by a time step $\delta t$ as:
  

\begin{eqnarray}
&& \exp [- i  \widetilde{H}  \delta t ]  \approx  \\ \nonumber
&& \exp[- i H_{\Delta} \delta t/2]  \exp[- i H_{\rm{KE}} \delta t] \exp[- i H_{\Delta} \delta t/2] + O(\delta t^3)
\end{eqnarray}
where 

\begin{eqnarray} 
&& H_{\rm{KE}} = \\ \nonumber
&& \left( \begin{array}{ccc}
-\frac{\hbar^2}{2m}(  \frac{\partial^2}{\partial x^2}+ \frac{\partial^2}{\partial y^2} - k_z^2)- \mu & 0  \\
0 & \frac{\hbar^2}{2m}(\frac{\partial^2}{\partial x^2}+ \frac{\partial^2}{\partial y^2} - k_z^2)  + \mu \end{array} \right)
 \end{eqnarray}

and 
\begin{equation}
H_{\rm{\Delta}} = \left( \begin{array}{ccc}
0 & \Delta(x,y, t)   \\
\Delta^*(x,y, t)  & 0  \end{array}\right)
\end{equation}

Because $H_{\rm{\Delta}} $ is diagonal in real space, applying the operator $\exp(-i H_{\Delta} \delta t/2)$ to $\left( \begin{array}{ccc}
u_{n, k_z}(x, y, t)  \\
v_{n, k_z}(x,y, t) \end{array} \right) $ can be done with $O(N)$ operations, where $N$ is the number of grid points in the $x-y$ plane. Similarly, because $H_{\rm{KE}} $ is diagonal in $k$-space, the operator $\exp(-i H_{\rm{KE}} \delta t)$ can be applied to the Fourier transform of the $u$'s and $v$'s in $O(N)$ operations. Thus, our split-step algorithm can is represented formally as
\begin{widetext}
\begin{equation}
 \left( \begin{array}{ccc}
u_{n, k_z}(t+\delta t)  \\
v_{n, k_z}(t+ \delta t ) \end{array} \right) =  \exp[- i H_{\Delta} \delta t/2] \mathcal{F}^{-1} \exp[- i H_{\rm{KE}} \delta t] \mathcal{F} \exp[- i H_{\Delta} \delta t/2]  \left( \begin{array}{ccc}
u_{n, k_z}(t)  \\
v_{n, k_z}(t ) \end{array} \right) 
\end{equation}
\end{widetext}

where $\mathcal{F}$ represents the application of a Fast Fourier Transform which takes $O(N \log N)$. Note that in each time step we reevaluate $\Delta(x,y,t)$ using Eq.~(\ref{tddel}) with the updated $u$'s and $v$'s.


\end{document}